\documentclass[showpacs,preprint,preprintnumbers,amsmath,amssymb]{revtex4}
\usepackage{graphicx}
\usepackage{dcolumn}
\usepackage{bm}
\usepackage{epsfig}
\begin{document}

\preprint{}

\title{The Bifurcation
Theory of Magnetic Monopoles in a Charged Two-Condensate
Bose-Einstein System}

\author {Shu-Fan Mo}\thanks{Email: meshf07@lzu.cn.}
\author{Ji-Rong Ren}\thanks{Corresponding author. Email: renjr@lzu.edu.cn.}
\author {Tao Zhu}\thanks{Email: zhut05@lzu.cn.}

\affiliation{Institute of Theoretical Physics, Lanzhou University,
Lanzhou 730000, P. R. China}

\date{\today}

\begin{abstract}
Magnetic monopoles, that are particle-like field configurations with
which one can associate a topological charge, widely exist in
various three dimensional condensate systems. In this paper, by
making use of \emph{Duan}'s topological current theory, we obtain
the charge density of magnetic monopoles and their bifurcation
theory in a charged two condensate Bose-Einstein system. The
evolution of magnetic monopoles is studied from the topological
properties of a three-dimensional vector field. The magnetic
monopoles are found generating or annihilating at the limit points
and encountering, splitting, or merging at the bifurcation points.
\end{abstract}

\pacs{74. 20. De, 03. 75. Mn, 14. 80. Hv}


\maketitle
\section{introduction}
An elementary particle with a net magnetic charge is an old
hypothetical particle called magnetic monopole arise in classical
electromagnetism and has never been seen in real world. Modern
interest in magnetic monopole focus on quantum field theory, notably
Grand Unified Theories and superstring theories, that predict the
existence of possibility of magnetic monopoles. In 1931, Paul Dirac
\cite{dirac} proposed that the magnetic monopole with an attached
Dirac string may exist in quantum electrodynamics by their
phenomenon of electric charge quantization. In 1974, it was shown by
't Hooft \cite{thooft} and Polyakov \cite{Polyakov} that a magnetic
monopole could be regarded as topological excitations in a quantum
field theory due to the spontaneous symmetry breaking mechanism. The
quantized magnetic charge was interpreted as the topological charge
of the magnetic monopole. After 't Hooft and Polyakov's works, Duan
and Ge \cite{duange} studied the rigorous topological expressions of
many moving magnetic monopoles, which could not be derived from 't
Hooft and Polyakov's theory. It also revealed the inner structures
of the magnetic charge density current and showed that the zero
points of Higgs field were point-like source of magnetic monopole.
Recently, the theory of magnetic monopole has been frequently
employed in studying the Grand Unified Theories, the phase
transitions in the early universe, and the topological excitations
in condensed matter physics.

In condensed matter physics, there are also topological objects that
imitate magnetic monopoles. In chiral superconductor and superfluid,
the magnetic monopole excitations have been well studied by G. E.
Volovik \cite{cs}, and such magnetic monopole is the analog of Dirac
magnetic monopole which combined with two Abrikosov vortices or four
half-quantum vortices. These vortex lines represent the
``conventional" Dirac string. Such Dirac-like monopole has been
investigated also in ferromagnetic spinor Bose-Einstein
condensates\cite{use}. Beside the analog of Dirac magnetic monopole,
the `` 't Hooft-Polyakov monopole" can also be introduced to
condensed matter physics \cite{spinorbec,jiang}. In spinor
Bose-Einstein antiferromagnets, such point-like monopole has
recently been worked out by a number of authors
\cite{afbec,spinorbec}. Moreover, in charged two-condensate
Bose-Einstein system, such monopole which has a quantized magnetic
charge and can be regarded as a real magnetic monopole has been
proposed recently by Jiang \cite{jiang}. The induced magnetic field
of magnetic monopole and their rigorous density distribution
expression have been deduced by using \emph{Duan}'s topological
current theory \cite{mapping1,mapping2}. As indicated in above
paragraph, magnetic monopole excitations have already been studied
in the context of quantum field theory. Therefore, as pointed out in
Ref.\cite{afbec}, magnetic monopole excitations in condensed matter
offer the exciting opportunity to study the properties of magnetic
monopoles in detail. Undoubtedly, this will lead to important new
insights into the general topic of topological excitations in a
quantum field theory.

Furthermore, two-gap superconductivity has drawn great interest
recently due to the discovery of the two-band superconductor with
surprisingly high critical temperature MgB2 \cite{mb}. Two-gap
superconductivity is being supported by an increasing number of
experimental reports. Principally, the two gap superconductivity can
be investigated in the frame of a charged two-condensate Bose system
\cite{jiang,zhangxinhui,GLGP}. This system is described by a
Ginzburg-Landau model with two flavors of Cooper pairs.
Alternatively, it relates to a Gross-Pitaevskii functional with two
charged condensates of tightly bound fermion pairs, or some other
charged bosonic fields. Such theoretical models have a wider range
of applications, including interference between two Bose condensates
\cite{ts}, a multiband superconductor \cite{ts2}, two-component
Bose-Einstein condensates \cite{ts3}, and superconducting gap
structure of spin-triplet superconductor Sr2RuO4 \cite{ts4}. Using
this theoretical model, two typical topological excitations have
been presented. One is the knotted vortices \cite{GLGP,zhangxinhui},
and the other is the magnetic monopoles \cite{jiang}. The main
purpose of this paper is to discuss the topological properties of
the magnetic monopole excitations in charged two-condensate Bose
system.

In Ref.\cite{jiang}, Jiang has proposed the magnetic monopole
excitation in charged two-condensate Bose system, and by using the
\emph{Duan}'s topological current theory, the rigorous density
distribution expression of magnetic monopole has been deduced. The
topological charges of magnetic monopoles can be expressed in terms
of the Hopf indices and Brouwer degrees. However, Jiang's
conclusions base on a very important condition that the Jacobian
$D(\phi/x)\neq 0$ must be satisfied. When this condition fails, what
will happen? In this paper, we will investigate the behaviors of the
magnetic monopole when this condition fails.

This paper is arranged as follows. In Sec.II, we give a prime view
of the derivation of the topological structure of magnetic
monopoles. The magnetic monopoles are quantized at the topological
level and their quantum numbers are determined by the Hopf indices
and Brouwer degree. In Sec.III, we introduce the generation and
annihilation of magnetic monopoles at the limit point. The
bifurcation theory of magnetic monopoles at the first- and second-
order degenerate points are investigated in Section IV and V,
respectively. Section VI is our conclusions.
\section{Magnetic monopole excitations in charged two-condensate Bose-Einstein system: A prime introduction}
In order to make the background of this paper clear, in this section
we will give a brief review of the magnetic monopole excitations in
charged two-condensate Bose-Einstein system. Firstly, let us
consider a Bose-Einstein system with two electromagnetically
coupled, oppositely charged condensates, which can be described by a
two-flavor (denoted by $\alpha=1, 2$) Ginzburg-Landau or
Gross-Pitaevskii (GLGP) functional \cite{GLGP}, whose free energy
density is given by
\begin{eqnarray}
F=\frac{1}{2m_1}\big|(\hbar\partial_{\mu}+i\frac{2e}{c}A_{\mu})\Psi_1\big|^2
+\frac{1}{2m_2}\big|(\hbar\partial_{\mu}-i\frac{2e}{c}A_{\mu})\Psi_2\big|^2
+V(\Psi_{1,2})+\frac{\vec{B}^2}{8\pi},\label{landau}
\end{eqnarray}
in which
\begin{eqnarray}
V(\Psi_{1,2})=-b_{\alpha}|\Psi_{\alpha}|^2+\frac{c_{\alpha}}{2}|\Psi_{\alpha}|^4+
\eta[\Psi^{*}_1\Psi_2+\Psi^{*}_2\Psi_1],
\end{eqnarray}
where $\eta$ is a characteristic of interband Josephson coupling
strength \cite{strength}. The properties of the corresponding model
with a single charged two-condensate Bose-Einstein system are well
known. And the relevant field degrees of freedom are the massive
coefficient of the single complex order parameter and a vector field
that gains a mass because of the Meissner-Higgs effect. What is very
important in the present GLGP model is that the two charged fields
are not independent but nontrivially coupled through the
electromagnetic field, which indicate that there should be a
nontrivial, hidden topology in this system. However, it cannot be
recognized obviously in the form of Eq.(\ref{landau}). For working
out the topological structure and studying it conveniently, we need
to reform the GLGP functional. Ref.\cite{GLGP} introduces a set of
variables $\rho$ and $\chi_{1,2}$ by
\begin{eqnarray}
\Psi_\alpha=\sqrt{2m_\alpha}\rho\chi_\alpha,
\end{eqnarray}
where the complex $\chi_\alpha=|\chi_\alpha|e^{i\varphi_\alpha}$
satisfying $|\chi_1|^2+|\chi_2|^2=1$ and $\rho$ has the following
expression
\begin{eqnarray}
\rho^2=\frac{1}{2}(\frac{|\Psi_1|^2}{m_1}+\frac{|\Psi_2|^2}{m_2}).
\end{eqnarray}
Here $\rho$ is a massive field which is related to the densities of
the Cooper pair. Using the variables $\chi_{1,2}$ and Pauli matrices
$\sigma$, we define the three dimensional unit vector field
$\vec{n}=(\bar{\chi}, \vec{\sigma}\chi)$, where $(,)$ denotes the
scalar product and $\bar{\chi}=(\chi_1^*~\chi_2^*)$. Then the
original GLGP free energy density Eq.(\ref{landau}) can be
represented as
\begin{eqnarray}
F=\frac{\hbar^2\rho^2}{4}(\partial\vec{n})^2+\hbar^2(\partial\rho)^2+\frac{\rho^2}{16}\vec{C}^2+V(\rho,
n_1, n_3)\nonumber\\ +\frac{\hbar^2c^2}{512\pi
e^2}(\frac{1}{\hbar}[\partial_{\mu}C_{\nu}-\partial_{\nu}C_{\mu}]-\vec{n}\cdot\partial_{\mu}\vec{n}\times\partial_{\nu}\vec{n})^2,\label{chang}
\end{eqnarray}
where
\begin{eqnarray}
C_\mu=2i\hbar[\chi_1\partial_\mu\chi_1^*-\chi_1^*\partial_\mu\chi_1-
\chi_2\partial_\mu\chi_2^*+\chi_2^*\partial_\mu\chi_2]-\frac{8e}{c}A_\mu.
\end{eqnarray}
Now we find that there exists an exact equivalence between the
two-flavor GLGP mole and the nonlinear $O(3)~\sigma$ model
\cite{model} that is much more important to describe the topological
structure in high energy physics. In this paper, based on
\emph{Duan-Ge}'s decomposable gauge potential theory and
\emph{Duan}'s topological current theory, we display that there
exists another kind of topological defect, namely the magnetic
monopoles in this system.

As shown in Eq.(\ref{chang}), we know that the magnetic field of the
system can be divided into two parts. One part, the contribution of
field $C_{\mu}$ is introduced by the supercurrent density and can
only present us with the topological defects named vortices, as what
is in the single-condensate system. The other part is the
contribution
$\vec{n}\cdot\partial_{\mu}\vec{n}\times\partial_{\nu}\vec{n}$ to
the magnetic field, which originates from interactions of Cooper
pairs of two different flavors and is a fundamentally important
property of the two-condensate system.

The induced magnetic field $B_{\mu}$ due to
$\vec{n}\cdot\partial_{\mu}\vec{n}\times\partial_{\nu}\vec{n}$ term
is expressed as
\begin{eqnarray}
B^{\mu}=\frac{\hbar c}{8\pi
e}\varepsilon^{\mu\nu\lambda}\varepsilon_{abc}n^{a}\partial_{\nu}n^{b}\partial_{\lambda}n^{c}.
\end{eqnarray}
Then, the divergence of the induced magnetic field, namely Q, can be
represented in terms of the unit vector field $n^{a}$ as
\begin{eqnarray}
Q=\partial_{\mu}B^{\mu}=\frac{\hbar c}{8\pi
e}\varepsilon^{\mu\nu\lambda}\varepsilon_{abc}\partial_{\mu}n^{a}\partial_{\nu}n^{b}\partial_{\lambda}n^{c},
\end{eqnarray}
this is just the magnetic charge density of the system $\rho_{m}$,
which is the time component of the topological current
\begin{equation}
J^{\mu}_{m}=\frac{\hbar c}{8\pi
e}\epsilon^{\mu\nu\lambda\rho}\epsilon_{abc}\partial_{\nu}
n^{a}\partial_{\lambda}n^{b}\partial_{\rho}n^{c},~~~~~~(\mu, \nu,
\lambda, \rho = 0, 1, 2, 3).\label{current}
\end{equation}
It is easy to see that the current (\ref{current}) is identically
conserved,
\begin{eqnarray}
\partial_{\mu}J^{\mu}_{m}=0.
\end{eqnarray}
In order to investigate the topological structure of the magnetic
charge current, we introduce a three-component vector order
parameter $\vec{\phi}=(\phi^1,\phi^2,\phi^3)$ formed by the unit
vector $\vec{n}$ which satisfies
\begin{equation}
n^{a}=\frac{\phi^{a}}{\|\phi\|},~~~
\|\phi\|=\sqrt{\phi^{a}\phi^{a}},~~~~~~ (a=1,2,3).\label{vector}
\end{equation}
Obviously, the order parameter $\vec{\phi}$ can be looked upon as a
smooth mapping between the three-dimensional space $X$ (with the
local coordinate $x$) and the three-dimensional Euclidean space
$R^3$ $\phi:x\longmapsto\vec{\phi}(x)\in R^3$. $n^{a}$ is a section
of sphere bundle $S(X)$.

Applying \emph{Duan}'s topological current theory
\cite{mapping1,mapping2}, one can obtain
\begin{equation}
J^{\mu}_{m}=\frac{\hbar
c}{e}\delta^3(\vec{\phi})D^{\mu}(\frac{\phi}{x}),\label{delt}
\end{equation}
and the Jacobian $D^{\mu}(\frac{\phi}{x})$ is defined as
\begin{equation}
\epsilon^{abc}D^{\mu}(\frac{\phi}{x})=\epsilon^{\mu\nu\lambda\rho}\partial_{\nu}
\phi^{a}\partial_{\lambda}
\phi^b\partial_{\rho}\phi^{c}.\label{Jacobian}
\end{equation}
The delta function expression (\ref{delt}) of the topological
current $J^{\mu}_{m}$ tells us that it does not vanish only at the
zero points of $\vec{\phi}$, i.e., the sites of the magnetic
monopole. The implicit function theorem \cite{imp} shows that under
the regular condition
\begin{eqnarray}
D^0(\frac{\phi}{x})\neq 0,
\end{eqnarray}
the general solutions of
\begin{eqnarray}
\phi^{a}(x^{1},x^{2},x^{3},t)=0,~~~~~~(a=1,2,3).\label{aa}
\end{eqnarray}
The solutions of Eq.(\ref{aa}) can be generally expressed as
$$x^1=x^1_{i}(t), x^2=x^2_{i}(t),
x^3=x^3_{i}(t),~~~~~~(i=1,2,\cdots,K)$$ that represent the world
lines of $K$ isolated zero points $\vec{z_{i}}(t)(i=1,2,\cdots,K)$.
These zero points are just the magnetic monopole excitations, and
the $ith$ world line $\vec{z_{i}}(t)$ determines the motion of the
$i$th magnetic monopole.

The $\delta-$ function theory \cite{delta} demonstrates the relation
$$\delta^3(\vec{\phi})=\sum^{K}_{i=1}\frac{\beta_{i}}{|D(\frac{\phi}{x})|_{\vec{z_{i}}}}\delta^3(\vec{r}-z_{i}(t)),$$
where the positive integer $\beta_{i}$ is the Hopf index of $\phi-$
mapping, which means that when $\vec{r}$ covers the neighborhood of
the zero point $\vec{z_{i}}(t)$ once, the vector field $\vec{\phi}$
covers the corresponding region in $\phi$ space $\beta_{i}$ times,
which is a topological number of first Chern class and relates to
the generalized winding number of the $\phi-$ mapping. With the
definition of the vector Jacobian (\ref{Jacobian}), and using the
implicit function theorem, the general velocity of the $ith$
magnetic monopole can be introduced
\begin{eqnarray}
V^{\mu}_{i}=\frac{dz^{\mu}_{i}}{dt}=\left.\frac{D^{\mu}(\frac{\phi}{x})}{D(\frac{\phi}{x})}\right|_{\vec{z_{i}}},~~~
V^0_{i}=1.\label{velocity}
\end{eqnarray}
Then, we can get the magnetic charge current $J^{\mu}_{m}$ in the
form of the current and the density of a system of $K$ classical
point particles in $(3+1)-$ dimensional space-time with topological
charge $W_{i}=\beta_{i}\eta_{i}$ :
\begin{eqnarray}
\vec{j_{m}}&=&\frac{\hbar c}{e}\sum^{K}_{i=1}W_{i}\vec{V_{i}}\delta^3(\vec{r}-\vec{z_{i}}(t)),\nonumber\\
\rho_{m}&=&\frac{\hbar
c}{e}\delta^3(\vec{\phi})D(\frac{\phi}{x})=\frac{\hbar
c}{e}\sum^{K}_{i=1}W_{i}\delta^3(\vec{r}-\vec{z_{i}}(t)),\label{motion}
\end{eqnarray}
where $\eta_{i}=sgn(D(\frac{\phi}{x})|_{\vec{z_{i}}})=\pm 1$ is the
Brouwer degree, and $W_{i}=\beta_{i}\eta_{i} $ is the winding number
of $\vec{\phi}$ at the zero point $\vec{z_{i}}(t)$. It is clear that
Eq.(\ref{motion}) describes the motion of the magnetic monopoles in
space-time, and the topological quantum numbers are determined by
the Hopf indices $\beta_{i}$ and Brouwer degrees $\eta_{i}$ of the
$\phi-$ mapping at its zeros. Here, $\eta_{i}=+1$ corresponds to a
magnetic monopole and $\eta_{i}=-1$ corresponds to an anti-magnetic
monopole.
\section{The generation and annihilation of magnetic monopoles}
As investigated before, the equations of $\vec{\phi}$'s zeros play
an important role in describing the topological structures of the
magnetic monopole in charged two-condensate Bose-Einstein system.
Now we begin discussing the properties of the zero points, in other
words, the properties of the solutions of Eq.(\ref{aa}). As we knew
before, if the Jacobian
\begin{equation}
D^0(\frac{\phi}{x}) \neq 0,\label{reu}
\end{equation}
we will have the isolated zeros of the vector field $\vec{\phi}$.
The isolated solutions are called regular points. However, when the
condition (\ref{reu}) fails, the usual implicit function theorem
\cite{imp} is of no use. The above discussion will change in some
way and lead to the branch process. Now, we denote one of the zero
points as $(t^*,\vec{x^*})$. Let us explore what happen to the
magnetic monopoles. In \emph{Duan}'s topological current theory,
there are usually two kinds of branch points, the limit points and
bifurcation points, satisfying
\begin{equation}
\left.D^{i}(\frac{\phi}{x})\right|_{(t^*,\vec{x^*})} \neq
0,~~~~~~i=1,2,3\label{limit}
\end{equation}
and
\begin{equation}
\left.D^{i}(\frac{\phi}{x})\right|_{(t^*,\vec{x^*})}=0,~~~~~~i=1,2,3,\label{bifurcation}
\end{equation}
respectively. Here, we consider the case (\ref{limit}). The other
case (\ref{bifurcation}) is complicated and will be treated in
Section 3 and 4. In order to be simple and without lose generality,
we choose $i=1$.

If the Jacobian
\begin{equation}
\left.D^1(\frac{\phi}{x})\right|_{(t^*,\vec{x^*})}\neq 0,
\end{equation}
we can use the Jacobian $D^1(\frac{\phi}{x})$ instead of
$D^0(\frac{\phi}{x})$ for the purpose of using the implicit function
theorem. This means we will replace the timelike variable $x^0=t$ by
$x^1$. For seeing this point clearly, we rewrite the equations of
(\ref{aa}) as
\begin{equation}
\vec{\phi}(x^1,x^2,x^3,t)=0.
\end{equation}
Then we have a unique solution of Eq.(\ref{aa}) in the neighborhood
of the limit point $(t^*,\vec{x^*})$
\begin{equation}
t=t(x^1),~~~~~x^2=x^2(x^1),~~~~~~x^3=x^3(x^1),
\end{equation}
with $t^*=t(x^{1*})$. We call the critical points $(t^*,\vec{x^*})$
the limit points. In the present case, we know that
\begin{equation}
\left.\frac{dx^1}{dt}\right|_{(t^*,\vec{x^*})}=\left.\frac{D^1(\frac{\phi}{x})}{D(\frac{\phi}{x})}\right|_{(t^*,\vec{x^*})}=\infty
\label{wuqiong},
\end{equation}
i.e.,
\begin{equation}
\left.\frac{dt}{dx^1}\right|_{(t^*,\vec{x^*})}=0.
\end{equation}
Then the Taylor expansion of $t=t(x^1)$ at the limit point
$(t^*,\vec{x^*})$ is
\begin{equation}
t-t^*=\frac{1}{2}\left.\frac{d^2
t}{(dx^1)^2}\right|_{(t^*,\vec{x^*})}(x^1-z^1_l)^2,\label{cha}
\end{equation}
which is a parabola in $x^1-t$ plane. From Eq.(\ref{cha}) we can
obtain two solutions $x^1_1(t)$ and $x^1_2(t)$, which give two
branch solutions (world lines of magnetic monopoles). If
\begin{equation}
\left.\frac{d^2 t}{(dx^1)^2}\right|_{(t^*,\vec{x^*})}>0.
\end{equation}
We have the branch solutions for $t>t^*$ [see Fig.1(a)]; otherwise,
we have the branch solutions for $t<t^*$ [see Fig.1(b)]. These two
cases are related to the origin and annihilation of magnetic
monopoles.

\begin{figure}[h]
\begin{center}
\begin{tabular}{p{5cm}p{5cm}}
(a)\psfig{figure=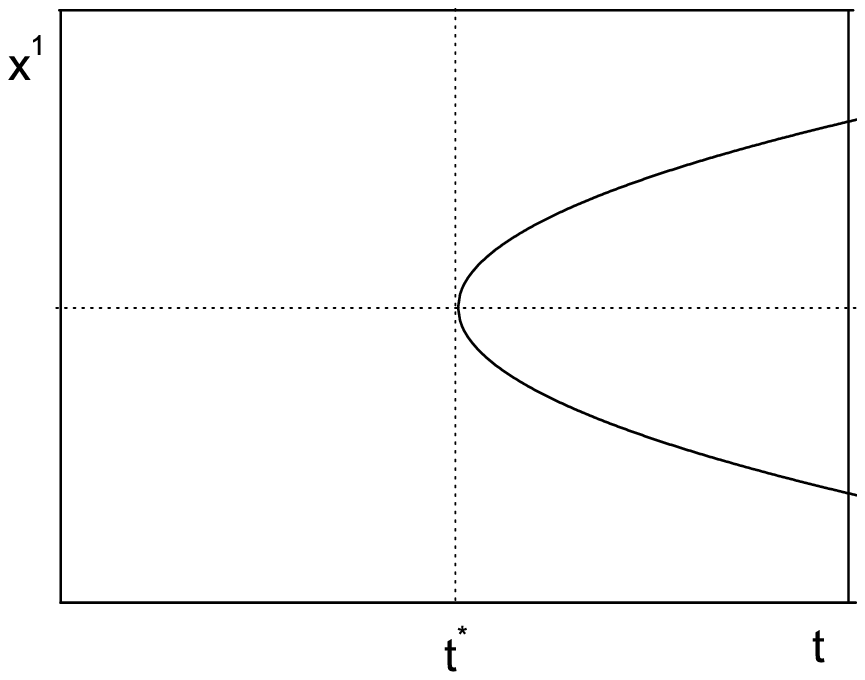,height=5cm,width=5cm}\\
(b)\psfig{figure=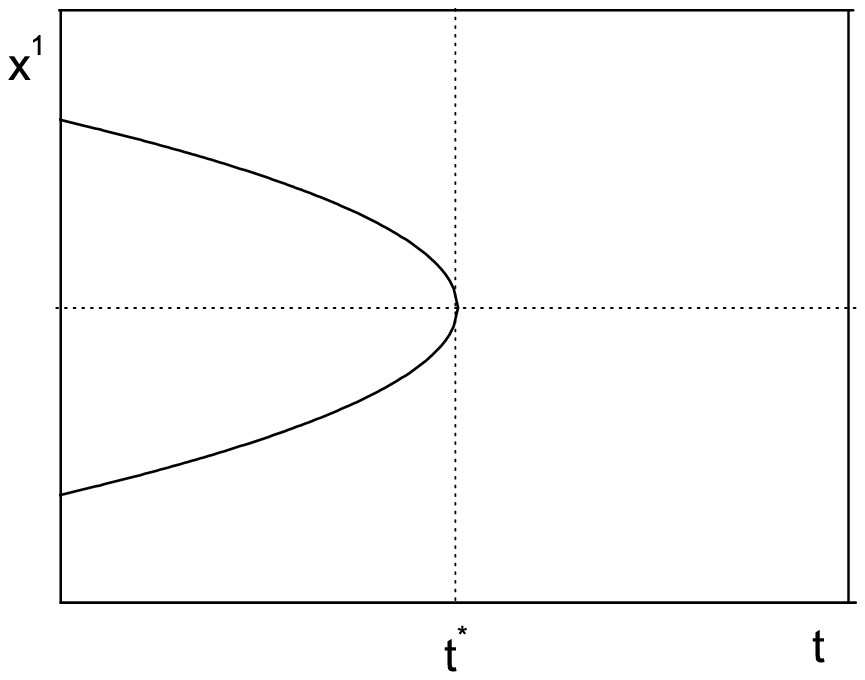,height=5cm,width=5cm}\\
\end{tabular}
\end{center}
\caption{Projecting the world lines of magnetic monopoles onto
$(x^1-t)$ plane. (a) The branch solutions for Eq.(\ref{cha}) when
$d^2t/(dx^1)^2|_{(t^*,\vec{z}_l)}>0$, i.e., two magnetic monopoles
with opposite charges generate at the limit point, i.e., the origin
of magnetic monopoles. (b) The branch solutions for Eq.(\ref{cha})
when $d^2t/(dx^1)^2|_{(t^*,\vec{z}_l)}<0$, i.e., two magnetic
monopoles with opposite charges annihilate at the limit point.}
\label{fig1}
\end{figure}

One of the results of Eq.(\ref{wuqiong}), that the velocity are
infinite when they are annihilating, agrees with the fact obtained
by Bray \cite{bray} who has a scaling argument associated with the
point defects final annihilation which leases to a large velocity
tail. From Eq.(\ref{wuqiong}), we also obtain a new result that the
velocity field is infinite when they are generating, which is gained
only from the topology of the vector function $\vec{\phi}$.

Since topological current is identically conserved, the topological
charges of these two generated or annihilated magnetic monopoles
must be opposite at the limit point, i.e.,
\begin{equation}
\beta_{l_1}\eta_{l_1}=-\beta_{l_2}\eta_{l_2},\label{cc1}
\end{equation}
which shows that $\beta_{l_1}=\beta_{l_2}$ and
$\eta_{l_1}=-\eta_{l_2}$, which is important in the charged
two-component Bose-Einstein system. One can see the fact that the
Brouwer degree $\eta$ is indefinite at the limit points implies and
can change discontinuously at limit points along the world lines of
the magnetic monopoles (from $\pm1$ to $\mp1$).

For a limit point it is required that
$D^1(\frac{\phi}{x})\big|_{(t^*,\vec{x^*})}\neq 0$. As to a
bifurcation point \cite{bif}, it must satisfy a more complex
condition. This case will be discussed in the following section.

\section{Bifurcation of magnetic monopoles}

In this section we have the restrictions of Eq.(\ref{bifurcation})
at the bifurcation points $(t^{*},\vec{x^*})$,
\begin{equation}
\left.D(\frac{\phi}{x})\right|_{(t^*,\vec{x^*})}=0,~~~~\left.D^{i}(\frac{\phi}{x})\right|_{(t^*,\vec{x^*})}=0,~~~~~~i=1,2,3,\label{restrict}
\end{equation}
which leads to an important fact that the function relationship
between $t$ and  $\vec{x}$ is not unique in the neighborhood of the
bifurcation point $(t^{*}, \vec{x^*})$. In our dynamic form of
charge current, this fact can be seen easily from equation
(\ref{velocity})
\begin{equation}
\frac{dx^{i}}{dt}=\left.\frac{D^{i}(\frac{\phi}{x})}{D(\frac{\phi}{x})}\right|_{(t^*,\vec{x^*})},~~~~~~i=1,2,3,\label{velocity1}
\end{equation}
which under Eq.(\ref{restrict}) directly shows that the direction of
the integral curve of Eq.(\ref{velocity1}) is indefinite at $(t^{*},
\vec{x^*})$, i.e., the velocity field of the magnetic monopoles is
indefinite at $(t^{*},\vec{x^*})$. That is why the very point
$(t^{*}, \vec{x^*})$ is called a bifurcation point.

Assume that the bifurcation point $(t^{*}, \vec{x^*})$ has been
found from Eq.(\ref{aa}) and (\ref{restrict}). We know that, at the
bifurcation point, the rank of the Jacobian matrix
$[\frac{\partial\phi}{\partial x}]$ is less than 3. We suppose
\begin{equation}
\left.rank[\frac{\partial\phi}{\partial
x}]\right|_{(t^*,\vec{x^*})}=3-1=2,
\end{equation}
and let
\begin{equation}
\left.D^{i}(\frac{\phi}{x})\right|_{(t^*,\vec{x^*})}= \left|
  \begin{array}{cc}
    \frac{\partial\phi^1}{\partial x^2} & \frac{\partial\phi^1}{\partial
x^3} \\
    \frac{\partial\phi^2}{\partial x^2} & \frac{\partial\phi^2}{\partial
x^3} \\
  \end{array}
\right|_{\vec{x^*}}\neq 0,\label{matric}
\end{equation}
which means $x^*$ is a first-order degenerate point of the $\phi-$
mapping theory. (The case that $x^*$ is a second-order degenerate
point will be detailed in the next section.) From $\phi^1=0$ and
$\phi^2=0$, the implicit function theorem implies that there exists
one and only one system of function relationships
\begin{equation}
x^2=x^2(t,x^1),~~~~~~x^3=x^3(t,x^1).\label{relation}
\end{equation}
Substituting (\ref{relation}) into $\phi^1$ and $\phi^2$, we can
obtain
\begin{equation}
\phi^{b}(t, x^1, x^2(t, x^1), x^3(t, x^1))\equiv 0,~~~~~~~b=1,2
\end{equation}
which give
\begin{equation}
\sum^3_{j=2}\phi^{b}_{j}x^{j}_0=-\phi^{b}_0,~~~~~~\sum^3_{j=2}\phi^{b}_{j}x^{j}_1=-\phi^{b}_1,\label{function1}
\end{equation}
\begin{equation}
\sum^3_{j=2}\phi^{b}_{j}x^{j}_{00}=-\sum^3_{j=2}[2\phi^{b}_{j0}x^{j}_1+\sum^3_{k=2}(\phi^{b}_{jk}x^{k}_0)x^{j}_1]-\phi^{b}_{01},
\end{equation}
\begin{equation}
\sum^3_{j=2}\phi^{b}_{j}x^{j}_{01}=-\sum^3_{j=2}[\phi^{b}_{j0}x^{j}_1+\phi^{b}_{j1}x^{j}_0+\sum^3_{k=2}(\phi^{b}_{jk}x^{k}_0)x^{j}_0]-\phi^{b}_{00},
\end{equation}
\begin{equation}
\sum^3_{j=2}\phi^{b}_{j}x^{j}_{11}=-\sum^3_{j=2}[2\phi^{b}_{j1}x^{j}_1+\sum^3_{k=2}(\phi^{b}_{jk}x^{k}_1)x^{j}_1]-\phi^{b}_{11},
\end{equation}
where $b=1,2;~~~~~~j,k=2,3;$ and
\begin{eqnarray}
x^{j}_0=\frac{\partial x^{j}}{\partial t},~~~x^{j}_1=\frac{\partial
x^{j}}{\partial x^1},~~~x^{j}_{00}=\frac{\partial^2 x^{j}}{\partial
t^2},~~~ x^{j}_{01}=\frac{\partial^2 x^{j}}{\partial t\partial
x^1},~~~~~x^{j}_{11}=\frac{\partial^2 x^{j}}{(\partial x^1)^2},
\end{eqnarray}
\begin{eqnarray}
\phi^{b}_0=\frac{\partial\phi^{b}}{\partial
t},~~~\phi^{b}_1=\frac{\partial\phi^{b}}{\partial
x^1},~~~\phi^{b}_{j}=\frac{\partial\phi^{b}}{\partial x^{j}},~~~
\phi^{b}_{00}=\frac{\partial^2\phi^{b}}{\partial
t^2},~~~~~\phi^{b}_{01}=\frac{\partial^2\phi^{b}}{\partial t\partial
x^1},
\end{eqnarray}
\begin{eqnarray}
\phi^{b}_{11}=\frac{\partial^2\phi^{b}}{(\partial
x^1)^2},~~~\phi^{b}_{j0}=\frac{\partial^2\phi^{b}}{\partial
t\partial x^{j}},~~~
\phi^{b}_{j1}=\frac{\partial^2\phi^{b}}{\partial x^1\partial
x^{j}},~~~~~\phi^{b}_{jk}=\frac{\partial^2\phi^{b}}{\partial
x^{j}\partial x^{k}}.
\end{eqnarray}
From these expressions we can calculate the values of the first and
second order partial derivatives of (\ref{relation}) with respect to
$t$ and $x^1$ at the bifurcation point $\vec{x^*}$.

Here we must note that the above discussions do not relate to the
last component $\phi^3(\vec{x},t)$ of the vector order parameter
$\vec{\phi}$. With the aim of finding the different directions of
all branch curves at the bifurcation point, let us investigate the
Taylor expansion of
\begin{equation}
F(t, x^1)=\phi^3(t, x^1, x^2(t, x^1), x^3(t, x^1)),\label{Taylor}
\end{equation}
in the bifurcation point, which must vanish at the bifurcation
point, i.e.,
\begin{equation}
F(t^*, x^{1*})=0.
\end{equation}
From (\ref{Taylor}), the first-order partial derivatives of $F(t,
x^1)$ is
\begin{eqnarray}
\frac{\partial F }{\partial t}=\frac{\partial\phi^3}{\partial
t}+\sum^3_{j=2}\frac{\partial\phi^3}{\partial x^{j}}x^{j}_0,~~~
\frac{\partial F }{\partial x^1}=\frac{\partial\phi^3}{\partial
x^1}+\sum^3_{j=2}\frac{\partial\phi^3}{\partial
x^{j}}x^{j}_1.\label{first}
\end{eqnarray}
On the other hand, making use of (\ref{matric}), (\ref{function1}),
(\ref{first}), and Cramer's rule, it is not difficult to prove that
the two restrictive conditions in (\ref{restrict}) can be rewritten
as
\begin{eqnarray}
\left.D(\frac{\phi}{x})\right|_{(t^*,\vec{x^*})}=\left.\frac{\partial
F}{\partial x^1}D^1(\frac{\phi}{x})\right|_{(t^*,\vec{x^*})}=0,
\end{eqnarray}
\begin{eqnarray}
\left.D^1(\frac{\phi}{x})\right|_{(t^*,\vec{x^*})}=\left.\frac{\partial
F}{\partial t}D^1(\frac{\phi}{x})\right|_{(t^*,\vec{x^*})}=0.
\end{eqnarray}
By considering (\ref{matric}), the above equations give
\begin{eqnarray}
\left.\frac{\partial F}{\partial
t}\right|_{(t^*,\vec{x^*})}=0,~~~~~\left.\frac{\partial F}{\partial
x^1}\right|_{(t^*,\vec{x^*})}=0.\label{lead to}
\end{eqnarray}
The second-order partial derivatives of the function $F(t,x^1)$ are
easily found to be
\begin{eqnarray}
\frac{\partial^2 F}{\partial
t^2}=\phi^3_{00}+\sum^3_{j=2}[2\phi^3_{j0}x^{j}_0+\phi^3_{j}x^{j}_{00}
+\sum^3_{k=2}(\phi^3_{jk}x^{k}_0)x^{j}_0],
\end{eqnarray}
\begin{eqnarray}
\frac{\partial^2 F}{\partial t\partial
x^1}=\phi^3_{11}+\sum^3_{j=2}[\phi^3_{j0}x^{j}_1+\phi^3_{j1}x^{j}_0+\phi^3_{j}x^{j}_{01}
+\sum^3_{k=2}(\phi^3_{jk}x^{k}_0)x^{j}_0],
\end{eqnarray}
\begin{eqnarray}
\frac{\partial^2 F}{(\partial
x^1)^2}=\phi^3_{11}+\sum^3_{j=2}[2\phi^3_{j1}x^{j}_1+\phi^3_{j}x^{j}_{11}
+\sum^3_{k=2}(\phi^3_{jk}x^{k}_1)x^{j}_1],
\end{eqnarray}
which at $x^*=(t^*, \vec{x^*})$ are denoted by
\begin{eqnarray}
A=\left.\frac{\partial^2 F}{\partial
t^2}\right|_{(t^*,\vec{x^*})},~~~B=\left.\frac{\partial^2
F}{\partial t\partial
x^1}\right|_{(t^*,\vec{x^*})},~~~C=\left.\frac{\partial^2
F}{(\partial x^1)^2}\right|_{(t^*,\vec{x^*})},
\end{eqnarray}
where $j,k=2,3$ and
\begin{eqnarray}
\phi^3_{j}=\frac{\partial\phi^3}{\partial
x^{j}},~~~\phi^3_{00}=\frac{\partial^2\phi^3}{\partial
t^2},~~~\phi^3_{01}=\frac{\partial^2\phi^3}{\partial t\partial
x^1},~~~\phi^3_{11}=\frac{\partial^2\phi^3}{(\partial x^1)^2},
\end{eqnarray}
\begin{eqnarray}
\phi^3_{j0}&=&\frac{\partial^2\phi^3}{\partial t\partial
x^{j}},~~~\phi^3_{j1}=\frac{\partial^2\phi^3}{\partial x^1\partial
x^{j}},~~~\phi^3_{jk}=\frac{\partial^2\phi^3}{\partial x^{j}\partial
x^{k}}.
\end{eqnarray}
According to the \emph{Duan}'s topological current theory, the
Taylor expansion of the solution of $\phi^3$ in the neighborhood of
the bifurcation point can generally be denoted as
\begin{equation}
A(x^1-z^1_l)^2+2B(x^2-z^2_l)(t-t^*)+(t-t^*)^2=0,
\end{equation}
which is followed by
\begin{equation}
A(\frac{dx^1}{dt})^2+2B\frac{dx^1}{dt}+C=0,\label{shun}
\end{equation}
and
\begin{equation}
C(\frac{dt}{dx^1})^2+2B\frac{dt}{dx^1}+A=0,\label{dao}
\end{equation}
where $A$, $B$, and $C$ are three constants. The solutions of
Eq.(\ref{shun}) or Eq.(\ref{dao}) give different directions of the
branch curves (world lines of the magnetic monopoles) at the
bifurcation point. There are four kinds of important cases, which
will show the physical meanings of the bifurcation points.

\emph{Case 1($A\neq 0$)}. For $\Delta =4(B^2-AC)>0$, we get two
different directions of the velocity field of magnetic monopoles
\begin{equation}
\left.\frac{dx^1}{dt}\right|_{1,2}=\frac{-B\pm
\sqrt{B^2-AC}}{A},\label{case1}
\end{equation}
which are shown in Fig.2. It is the intersection of two magnetic
monopoles, which means that two magnetic monopoles meet and then
depart from each other at the bifurcation point.

\begin{figure}[h]
\begin{center}
\begin{tabular}{p{5cm}p{5cm}}
\psfig{figure=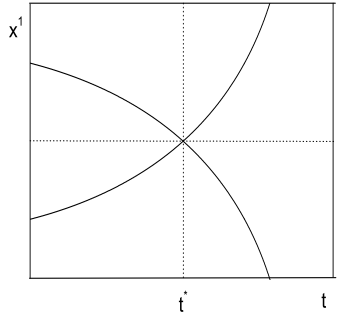,height=5cm,width=5cm}

\end{tabular}
\end{center}
\caption{Projecting the world lines of magnetic monopoles onto
$(x^1-t)$ plane. Two magnetic monopoles meet and then depart at the
bifurcation point.} \label{fig2}
\end{figure}

\emph{Case 2($A\neq 0$)}. For $\Delta =4(B^2-AC)=0$, the direction
of the velocity field of the magnetic monopole is only one
\begin{equation}
\left.\frac{dx^1}{dt}\right|_{1,2}=\frac{-B}{A},\label{case2}
\end{equation}
which includes three important situations. (a) One world line
resolves into two world lines, i.e., one magnetic monopole splits
into two magnetic monopoles at the bifurcation point [see Fig.3(a)].
(b) Two world lines merge into one magnetic monopole, i.e., two
magnetic monopoles merge into one magnetic monopole at the
bifurcation point [see Fig.3(b)]. (c) Two world lines tangentially
contact, i.e., two magnetic monopoles tangentially encounter at the
bifurcation point [see Fig.3(c)].

\begin{figure}[h]
\begin{center}
\begin{tabular}{p{5cm}p{5cm}p{5cm}}
(a)\psfig{figure=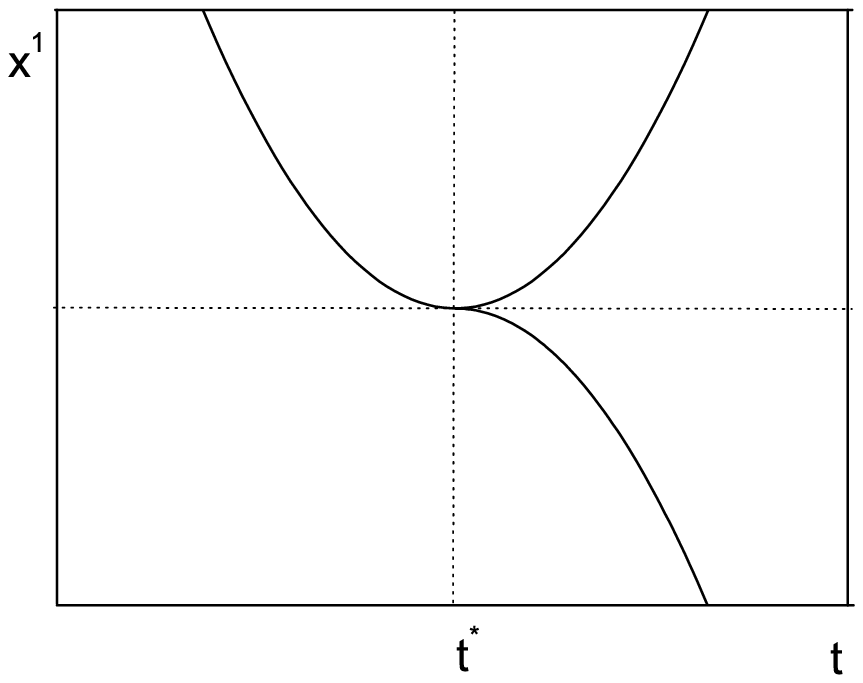,height=5cm,width=5cm} \\
(b)\psfig{figure=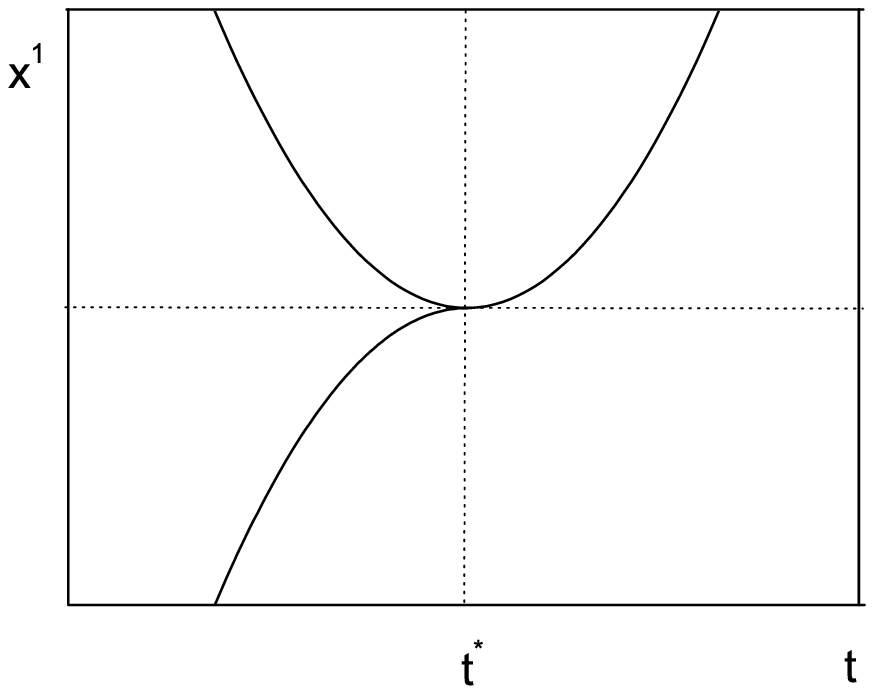,height=5cm,width=5cm} \\
(c)\psfig{figure=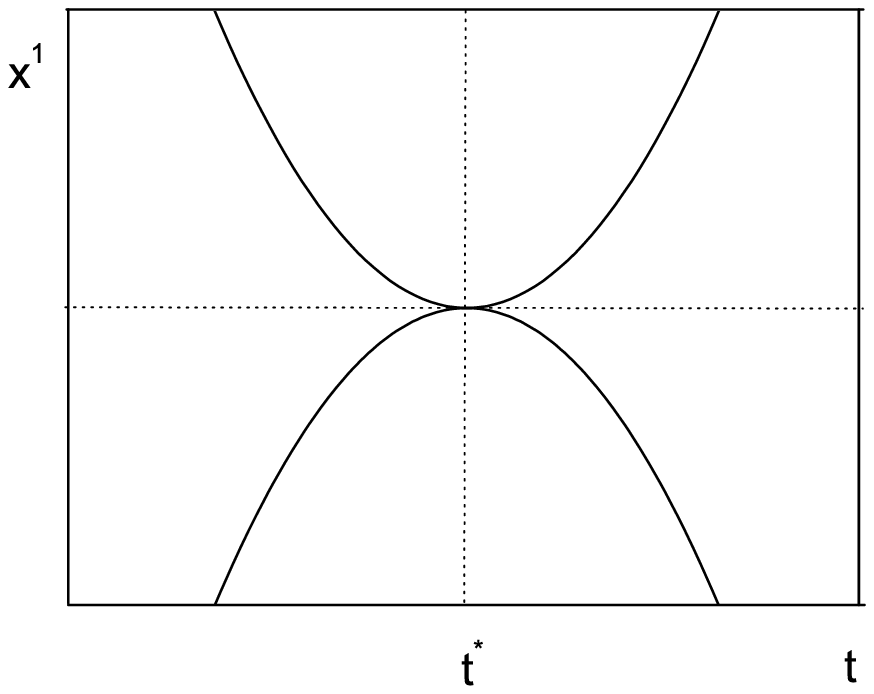,height=5cm,width=5cm}
\end{tabular}
\end{center}
 \caption{ (a) One magnetic monopole splits into two magnetic monopoles at the bifurcation point. (b) Two
magnetic monopoles merge into one magnetic monopole at the
bifurcation point. (c) Two world line of magnetic monopoles
tangentially intersect, i.e., two magnetic monopoles tangentially
encounter at the bifurcation point.} \label{fig3}
\end{figure}

\emph{Case 3($A=0, C\neq 0$)}. For $\Delta=4(B^2-AC)=0$, we have
\begin{equation}
\left.\frac{dt}{dx^1}\right|_{1,2}=\frac{-B\pm\sqrt{B^2-AC}}{C}=0,~~~-\frac{2B}{C}.\label{case3}
\end{equation}
There are two important cases: (a) Three world lines merge into one
world line, i.e., three magnetic monopoles merge into a magnetic
monopole at the bifurcation point [see Fig.4(a)]. (b) One world line
resolves into three world lines, i.e., a magnetic monopole splits
into three magnetic monopoles at the bifurcation point [see
Fig.4(b)].

\emph{Case 4($A=C=0$)}. Eq.(\ref{shun}) and Eq.(\ref{dao}) give
respectively
\begin{equation}
\frac{dx^1}{dt}=0,~~~\frac{dt}{dx^1}=0.\label{case4}
\end{equation}
This case is obvious [see Fig.5], and similar to Case 3.

The above solutions reveal the evolution of the magnetic monopoles.
Besides the encountering of the magnetic monopoles, i.e., two
magnetic monopole encounter and then depart at the bifurcation point
along different branch cures [see Fig.2 and Fig.3(c)], it also
includes splitting and merging of magnetic monopoles. When a
multi-charged magnetic monopole moves through the bifurcation point,
it may split into several magnetic monopoles along different branch
curves [see Fig.3(a), Fig.4(b), Fig.5(b)]. On the contrary, magnetic
monopoles can merge into a magnetic monopole at the bifurcation
point [see Fig.3(b) and Fig.4(a)].

\begin{figure}[h]
\begin{center}
\begin{tabular}{p{5cm}p{5cm}}
(a)\psfig{figure=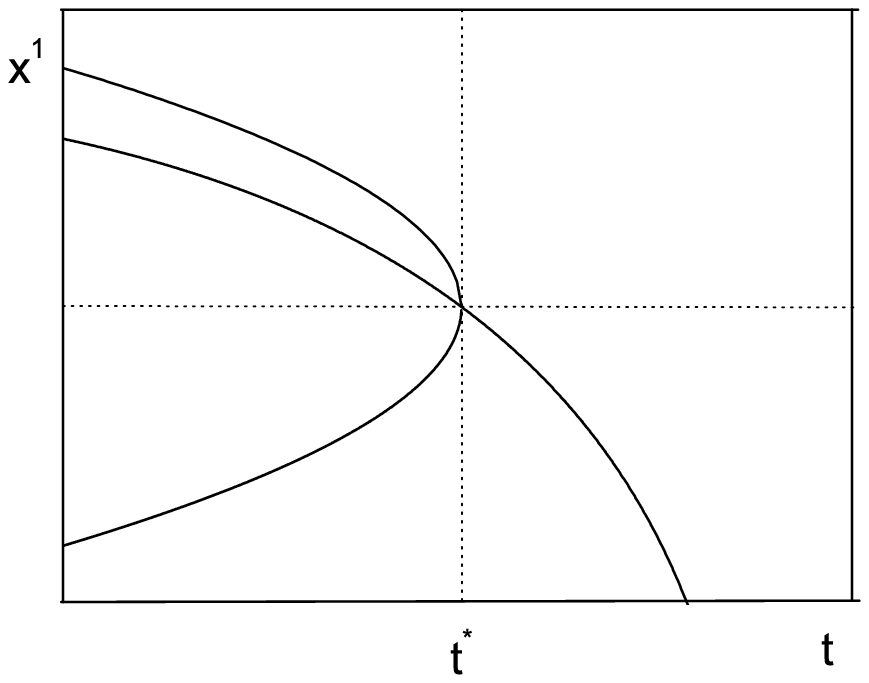,height=5cm,width=5cm} \\
(b)\psfig{figure=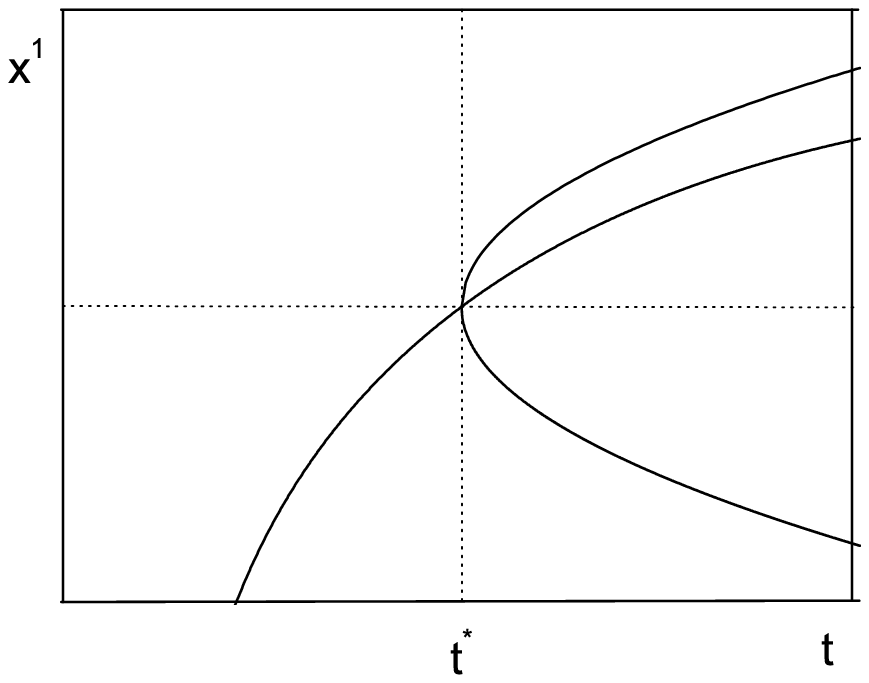,height=5cm,width=5cm}
\end{tabular}
\end{center}
\caption{ Two important cases of Eq.(\ref{case3}). (a) Three
magnetic monopoles merge into one at the bifurcation point. (b) One
magnetic monopole splits into three magnetic monopoles at the
bifurcation point.} \label{fig4}
\end{figure}

\begin{figure}[h]
\begin{center}
\begin{tabular}{p{5cm}p{5cm}}
(a)\psfig{figure=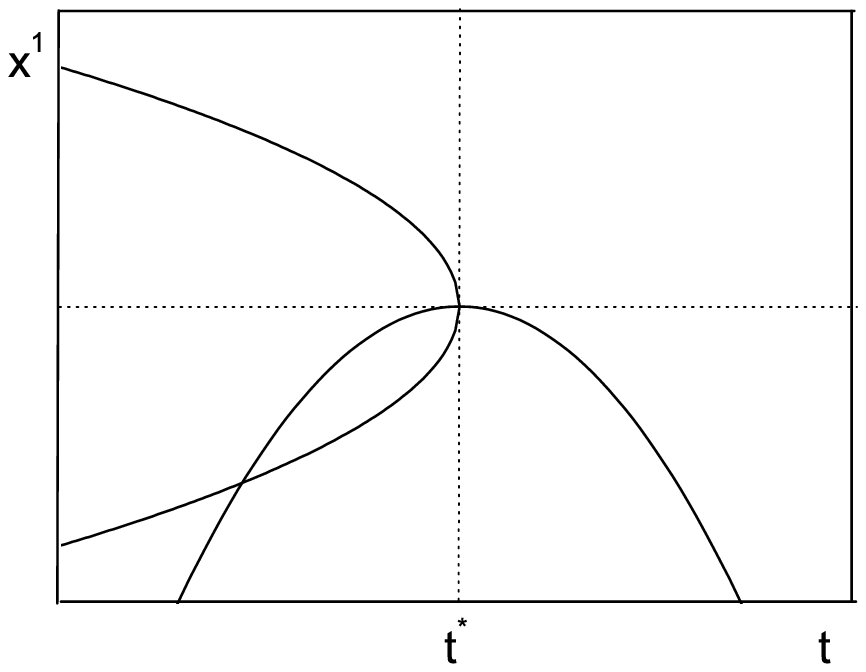,height=5cm,width=5cm} \\
(b)\psfig{figure=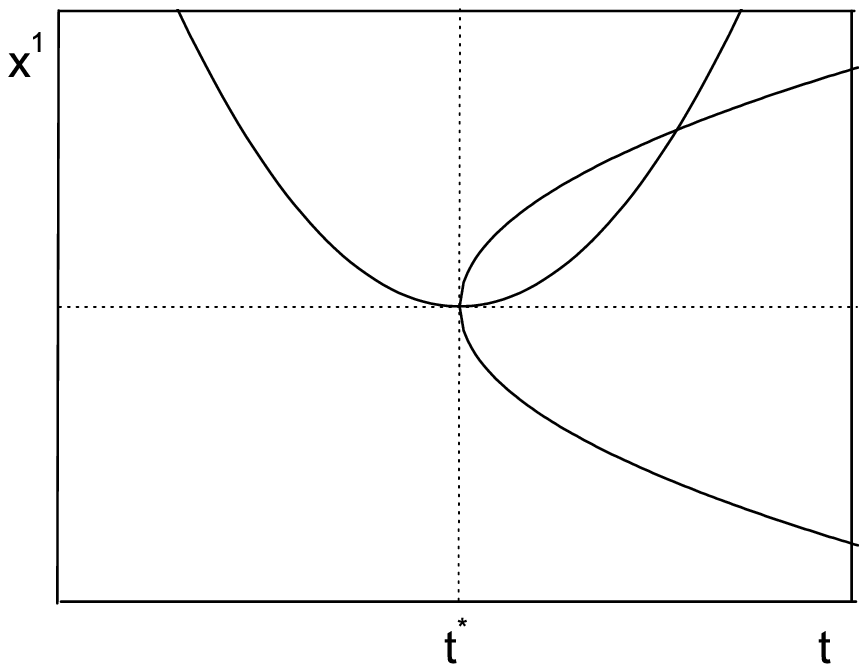,height=5cm,width=5cm}
\end{tabular}
\end{center}
\caption{ Two world lines intersect normally at the bifurcation
point. This case is similar to FIG.4. (a) Three magnetic monopoles
merge into one at the bifurcation point. (b) One magnetic monopole
splits into three magnetic monopoles at the bifurcation point.}
\label{fig5}
\end{figure}
At the same time, the remaining component can be deduced by
\begin{eqnarray}
\frac{dx^{j}}{dt}=x^{j}_0+x^{j}_1\frac{dx^1}{dt},~~~j=2,3.
\end{eqnarray}
As previous work, the identical conversation of the topological
charge shows the sum of the topological charge of these split
magnetic monopoles must be equal to that of the original magnetic
monopoles at the bifurcation point, i.e.,
\begin{equation}
\sum_{i}\beta_{l_{i}}\eta_{l_{i}}=\sum_{f}\beta_{l_{f}}\eta_{l_{f}},\label{cc2}
\end{equation}
for fixed $l$. Furthermore, from the above studies, we see that the
generation, annihilation, and bifurcation of magnetic monopoles are
not gradually changed, but suddenly changed at the critical points.
\section{the bifurcation of magnetic monopole at a second-order degenerate point}
In the preceding section we studied the bifurcation of a magnetic
monopole at a first-order degenerate point. In this section, we
investigate the branching process of the magnetic charge current at
a second-order degenerate point $x^*=(t^*, \vec{x^*})$, at which the
rank of the Jacobian matrix $[\frac{\partial\phi}{\partial x}]$ is
\begin{eqnarray}
\left.rank[\frac{\partial\phi}{\partial
x}]\right|_{(t^*,\vec{x^*})}=3-2=1.
\end{eqnarray}
Suppose that
\begin{eqnarray}
\left.\frac{\partial\phi^1}{\partial
x^3}\right|_{(t^*,\vec{x^*})}\neq 0.
\end{eqnarray}
With the same reasons as in obtaining (\ref{relation}), in the
neighborhood of $x^*$, from $\phi^1(x)=0$ we have the function
relationship
\begin{eqnarray}
x^3=x^3(t, x^1, x^2).\label{function2}
\end{eqnarray}
In order to determine the values of the first and second order
partial derivatives of $x^3$ with respect to $t, x^1,$ and $x^2$,
one can substitute the relationship (\ref{function2}) into
$\phi^2(x)=0$ and $\phi^3(x)=0$. Then, we get
\begin{eqnarray}
F_1(t, x^1, x^2)=\phi^2(t, x^1, x^2, x^3(t, x^1, x^2))=0,\nonumber\\
F_2(t, x^1, x^2)=\phi^3(t, x^1, x^2, x^3(t, x^1, x^2))=0.
\end{eqnarray}
For calculating the partial derivatives of the function $F_1$ and
$F_2$ with respect to $t, x^1,$ and $x^2$, one can take notice of
(\ref{function2}) and use six similar expressions to (\ref{lead
to}), i.e.,
\begin{eqnarray}
\left.\frac{\partial F_{c}}{\partial
t}\right|_{(t^*,\vec{x^*})}=0,~~~\left.\frac{\partial
F_{c}}{\partial
x^1}\right|_{(t^*,\vec{x^*})}=\phi,~~~\left.\frac{\partial
F_{c}}{\partial x^2}\right|_{(t^*,\vec{x^*})}=\phi,~~~c=1,2.
\end{eqnarray}
So the Taylor expansions of $F_1(t, x^1, x^2)$ and $F_2(t, x^1,
x^2)$ can be written in the neighborhood of $(t^*,\vec{x^*})$ by
\begin{eqnarray}
F_{c}(t, x^1, x^2)&\approx&
A_{c1}(t-t^*)^2+A_{c2}(t-t^*)(x^1-x^{1*})\nonumber\\
&+&A_{c3}(t-t^*)(x^2-x^{2*})+ A_{c4}(x^1-x^{1*})^2
+A_{c5}(x^1-x^{1*})(x^2-x^{2*})\nonumber\\
&+&A_{c6}(x^2-x^{2*})^2=0,\label{expansion}
\end{eqnarray}
where $c=1,2$ and
\begin{eqnarray}
A_{c1}&=&\frac{1}{2}\left.\frac{\partial^2 F_{c}}{\partial
t^2}\right|_{(t^*,\vec{x^*})},~~~A_{c2}=\left.\frac{\partial^2
F_{c}}{\partial t\partial
x^1}\right|_{(t^*,\vec{x^*})},~~~A_{c3}=\left.\frac{\partial^2
F_{c}}{\partial t\partial x^2}\right|_{(t^*,\vec{x^*})},\nonumber\\
A_{c4}&=&\frac{1}{2}\left.\frac{\partial^2 F_{c}}{(\partial
x^1)^2}\right|_{(t^*,\vec{x^*})},~~~A_{c5}=\left.\frac{\partial^2
F_{c}}{\partial x^1\partial x^2}\right|_{(t^*,\vec{x^*})},~~~
A_{c6}=\frac{1}{2}\left.\frac{\partial^2 F_{c}}{(\partial
x^2)^2}\right|_{(t^*,\vec{x^*})}.
\end{eqnarray}
Dividing (\ref{expansion}) by $(t-t^*)^2$ and taking the limit
$t\rightarrow t^*$, one obtains the two quadratic equations of
$\frac{dx^1}{dt}$ and $\frac{dx^2}{dt}$,
\begin{eqnarray}
A_{c1}+A_{c2}\frac{dx^1}{dt}+A_{c3}\frac{dx^2}{dt}+A_{c4}(\frac{dx^1}{dt})^2
+A_{c5}\frac{dx^1}{dt}\frac{dx^2}{dt}+A_{c6}(\frac{dx^2}{dt})^2=0,
\end{eqnarray}
and further, eliminating the variable $\frac{dx^1}{dt}$, one has the
equation of $\frac{dx^2}{dt}$ in the form of a determinant
\begin{eqnarray}
\left|
  \begin{array}{cccc}
    A_{14}~~~& A_{15}v+A_{12}~~~ & A_{16}v^2+A_{13}v+A_{11}~~~~ & 0 \\
    0 & A_{14} & A_{15}v+A_{12} & A_{16}v^2+A_{13}v+A_{11} \\
    A_{24} & A_{25}v+A_{22} &A_{26}v^2+A_{23}v+A_{21} & 0 \\
    0 & A_{14} & A_{25}v+a_{22} & A_{26}v^2+A_{23}v+A_{21} \\
  \end{array}
\right|=0,
\end{eqnarray}
with the variable $v=\frac{dx^2}{dt}$, which is a four-order
equation of $\frac{dx^2}{dt}$
\begin{eqnarray}
a_1(\frac{dx^2}{dt})^4+a_2(\frac{dx^2}{dt})^3+a_3(\frac{dx^2}{dt})^2+a_4(\frac{dx^2}{dt})+a_5=0.
\end{eqnarray}
Hence, different directions of the branch curves at the second-order
degenerate point $x^*$ is structured. The largest number of
different branch curves is four, which means an original magnetic
monopole with the topological quantum $\beta\eta$ can split into at
most four particles at one time with charges
$\beta_{l}\eta_{l}~(l=1, 2, 3, 4)$ satisfying
\begin{eqnarray}
\beta_{1}\eta_{1}+\beta_{2}\eta_{2}+\beta_{3}\eta_{3}+\beta_{4}\eta_{4}=\beta\eta.
\end{eqnarray}
\section{Conclusions}
Our conclusions can be summarized as follows: First, in charged
two-component Bose-Einstein system, we obtained the dynamic form of
magnetic monopole and quantized the magnetic charge at the
topological level in units of $\frac{\hbar c}{e}$. The topological
quantum numbers are determined by the Hopf indices and Brouwer
degrees (i.e. the winding numbers), which are topological numbers.
Second, the evolution of magnetic monopoles is studied from the
topological properties of a three-dimensional vector field
$\vec{\phi}$. We find that there exist crucial cases of branch
processes in the evolution of the magnetic monopoles when
$D(\frac{\phi}{x})\neq 0$, i.e., $\eta_l$ is indefinite. This means
that the magnetic monopoles generate or annihilate at the limit
points and encounter, split, or merge at the bifurcation points of
the three-dimensional vector field $\vec{\phi}$, which shows that
the magnetic monopoles system is unstable at these branch points.
Third, we show the result that the velocity of magnetic monopole is
infinite when they are annihilating or generating, which are
obtained only from the topological properties of the
three-dimensional vector field $\vec{\phi}$. Forth, we must point
out that there exist two restrictions of the evolution of magnetic
monopoles. One restriction is the conservation of the topological
charge of the magnetic monopoles during the branch process [see
Eq.(\ref{cc1}) and (\ref{cc2})], the other is that the number of
different directions of the world lines of magnetic monopoles is at
most $4$ at the bifurcation points [see Eq.(\ref{shun}) and
(\ref{dao})]. The first restriction is already known, but the second
is pointed out here for the first time to our knowledge. We hope
that it can be verified in the future. Finally, we would like to
point out that all the results in this paper have been obtained only
from the viewpoint of topology without using any particular models
or hypothesis.

\begin{acknowledgments}
Thanks to the works of Dr. X. H. Zhang and Dr. B. H. Gong in drawing
the figures in this paper. This work was supported by the National
Natural Science Foundation of China and the Cuiying Programme of
Lanzhou University.

\end{acknowledgments}

\end{document}